\documentstyle{article}[12pt]
\begin{document}
\begin{center}
{\large\bf UNIFIED FRACTIONAL KINETIC EQUATION AND 
                            A  FRACTIONAL DIFFUSION EQUATION}\\[1cm]
                           
R.K. SAXENA\\
Department of Mathematics and Statistics, Jai Narain Vyas University,\\  
Jodhpur - 342005, India\\[0.5cm]
A.M. MATHAI\\ 
Department of Mathematics and Statistics, McGill University,\\
Montreal, Canada H3A 2K6\\[0.5cm] 
H.J. HAUBOLD\\
Office for Outer Space Affairs, United Nations,\\
P.O. Box 500, A-1400 Vienna, Austria\\

\end{center}

{\bf Abstract.}  In earlier papers Saxena et al. (2002, 2003) derived the solutions of a number of fractional kinetic equations  in terms of generalized Mittag-Leffler functions  which extended the work of  Haubold and Mathai (2000). The object 
of the present paper is to investigate the solution of a unified form of fractional kinetic equation in which the free term contains any integrable function  $f(t)$, which provides 
the unification and extension of the results given earlier recently by Saxena et al.  (2002, 2003). The solution has been developed in terms of the Wright function  in a closed form by the method of Laplace transform. Further we derive a  closed-form solution of a fractional diffusion equation.  The asymptotic expansion of the derived  solution with respect to the space variable is also discussed. The results obtained  are in a form suitable for numerical computation.\\[0.5cm]
\section{Introduction}

Fundamental laws of physics are written as equations for the time evolution of a quantity $X(t)$, $dX(t)/dt = - AX(t)$, where this could be Maxwell's equations or Schroedinger's equation (if $A$ is limited to linear operators), or it could be Newton's law of motion or Einstein's equations for geodesics (if $A$ may also be a nonlinear operator). The mathematical solution (for linear operators) is $X(t) = X(0)\mbox{Exp}\left\{-At\right\}$. The initial value of the quantity at $t = 0$ is given by $X(0)$.

The same exponential behavior referred to above arises if $X(t)$ represents the scalar number density of species at time $t$ that do not interact with each other. If one denotes $A_p$ the production rate and $A_d$ the destruction rate, respectively, the number density $X(t)$ will obey an exponential equation where the coefficient $A$ is equal to the difference, $A_p - A_d$. Subsequently, $A_p^{-1}$ is the average time between production and $A_d^{-1}$ is the average time between destruction. This type of behavior arises frequently in biology, chemistry, and physics (Kaplan and Glass, 1998; Hilfer, 2000; Metzler and Klafter, 2000; Aslam Chaudhry and Zubair, 2002). This paper, in Section 3, considers the fractional generalization of the kinetic equation and derives closed form representation of its solution.

The evolution of the number density $X(x,t)$ relates the first derivative in time of this function to a spatial operator applied to the number density. The initial value of the function at time $t = 0$ is given by $X(x,0)$. In Section 4 of this paper, we give a relationship between the solution of the equation of evolution and the solution of the diffusion equation belonging to its fractional extension (Kaplan and Glass, 1998; Hilfer, 2000; Metzler and Klafter, 2000).

Section 2 summarizes mathematical results concerning solutions of the kinetic and diffusion equations in Sections 3 and 4, respectively, widely distributed in the literature or of very recent origin. These involve the Mittag-Leffler function, Wright function, and H-function, and the application of fractional calculus, Fourier transform, and Laplace transform to them.

Specifically, Haubold and Mathai (1995)  discussed a  conjecture of a variation of the solar neutrino signal looking at solutions of the standard kinetic equation based on lifetime densities and Poisson arrivals. A closed form representation of the fractional kinetic equation and thermonuclear function  is derived  by Haubold and Mathai (2000) in terms of certain series which represents the Mittag- Leffler function (Mittag-Leffler, 1903, 1905). In order to  extend the work of Haubold and Mathai (2000)  three fractional kinetic equations are solved by Saxena et al. (2002) in terms of the generalized Mittag-Leffler functions in a closed form.  Further extensions of these results are provided by  a recent paper of Saxena et al. (2003). Section 3 and 4 of this paper provide a unification and extension of the aforementioned results on fractional kinetic equations by investigating a closed-form solution of a unified fractional kinetic equation in which the free term contains a function f(t) by the method of Laplace transform. The solution has been obtained in terms of the generalized Wright function  in a   closed  form in Section 3. Also presented  is a closed-form solution of a fractional diffusion equation in terms of an H-function.  Its asymptotic expansion for large values of the space variable is discussed.  The results derived in this paper are in a form suitable for numerical computation.

\section{Mathematical Prerequisites}

      A generalization of the Mittag-Leffler function  (Mittag-Leffler, 1903, 1905)
\begin{equation}
      E_\alpha(z):=\sum^\infty_{n=0}\frac{z^n}{\Gamma(n\alpha+1)},\;\;(\alpha\in C, Re(\alpha)>0)
\end{equation}
was introduced by Wiman (1905) in the general  form 
\begin{equation}
     E_{\alpha,\beta}(z):= \sum^\infty_{n=0}\frac{z^n}{\Gamma(n\alpha+\beta)},\;\;(\alpha, \beta \in C, Re(\alpha)>0).
\end{equation}
The main results of these functions are available in the handbook of Erd\'elyi, Magnus, Oberhettinger and Tricomi (1955, Section 18.1) and the monographs written by Dzherbashyan (1966, 1993).
      Prabhakar  (1971) introduced a generalization of (2) in the form 
\begin{equation}      
E^\gamma_{\alpha,\beta}(z)=
\sum^\infty_{n=0}\frac{(\gamma)_nz^n}{\Gamma(n\alpha+\beta)(n)!},\;\;(\alpha, \beta, \gamma \in C, Re(\alpha)>0),
\end{equation}
where $(\gamma)_0=1$ is the Pochammer symbol, defined by 
\begin{equation}
(\gamma)_0=1,(\gamma)_k=\gamma(\gamma+1)\ldots(\gamma+k-1)\;\;(k=1,2,\ldots), \gamma\neq 0.
\end{equation}
It is an entire function with  $\rho=[Re(\alpha)]^{-1}$ (Prabhakar, 1971).\par
\noindent
For $\gamma=1$, this function coincides with (2), while for $\gamma=\beta=1$ with (1):
\begin{equation}
E^1_{\alpha,\beta}(z)=E_{\alpha,\beta}(z),\;\;E^1_{\alpha,1}(z)=E_\alpha(z).
\end{equation}
We also have
      \begin{equation}
 \Phi(\beta,\gamma,z)=\;_1F_1(\beta;\gamma;z)=\Gamma(\gamma)E_{1,\gamma}^\beta(z)                                     
\end{equation}
where  $\Phi(\beta,\gamma;z)$ is Kummer's confluent hypergeometric function  defined in Erd\'elyi et al. (1953, p.248, eq.1).

      The Mellin-Barnes integral representation for this function  follows from the integral 
\begin{equation}      
E^\gamma_{\alpha,\beta}(z)=\frac{1}{\Gamma(\gamma)}\frac{1}{2\pi\omega}\int_\Omega\frac{\Gamma(-\xi)\Gamma(\gamma+\xi)(-z)^\xi d\xi}{\Gamma(\beta+\xi \alpha)},
\end{equation}
where  $\omega=(-1)^{1/2}$. The contour $\Omega$ is a straight line parallel to the imaginary axis at a distance  `c' from the origin  and separating the poles of $\Gamma(-\xi)$  at the points $\xi=\nu \;\;(\nu=0,1,2,\ldots)$ from those of $\Gamma(\gamma+\xi)$ at the points $\xi=-\gamma-\nu\;\;(\nu=0,1,2,\ldots)$.
If we calculate the residues at the poles of $\Gamma(\gamma+\xi)$  at the points $\xi=-\gamma-\nu\;\;(\nu=0,1,2,\ldots),$ then it gives the analytic continuation formula of this function in the form 

\begin{equation}
E^\gamma_{\alpha,\beta}(z)=\frac{(-z)^{-\gamma}}{\Gamma(\gamma)}\sum^\infty_{\nu=0}\frac{\Gamma(\gamma+\nu)}{\Gamma[\beta-\alpha(\gamma+\nu)]}\frac{(-z)^{-\nu}}{(\nu)!},\mid z \mid>1.
\end{equation}                                         
 From (8) it follows that  for  large $z$  its behavior is given by 
 \begin{equation}
E^\gamma_{\alpha,\beta}(z) \sim O(\mid z\mid^{-\gamma}), \mid z\mid>1.
\end{equation}  
The H-function  is defined by means of a Mellin-Barnes type integral in the following manner  (Mathai and Saxena, 1978, p.2 ):
\begin{eqnarray}
H^{m,n}_{p,q}(z)& = & H^{m,n}_{p,q}\left[z\left|^{(a_p, A_p)}_{(b_q,B_q)}\right.\right] \nonumber\\           
&=&H^{m,n}_{p,q}\left[z \left|^{(a_1, A_1),\ldots,(a_p,A_p)}_{(b_1,B_1),\ldots,(b_q,B_q)}\right.\right]=\frac{1}{2\pi\omega}\int_\Omega\Theta(\xi)z^{-\xi} d\xi,
\end{eqnarray}
where
\begin{equation}
\Theta(\xi)=\frac{[\prod^m_{j=1}\Gamma(b_j+B_j\xi)][\prod^n_{j=1}\Gamma(1-a_j-A_j\xi)]}{[\prod^q_{j=m+1}\Gamma(1-b_j-B_j\xi)][\prod^p_{j=n+1}\Gamma(a_j+A_j\xi)]},
\end{equation}
$m,n,p,q\in N_0$ with $0\leq n\leq p,\;\; 1\leq m\leq q, \;\;A_i, B_j\in R_+,a_i,b_j \in R$ or $C(i=1,\ldots, p;\;\;j=1,\ldots, q)$ such that
\begin{equation}
A_i(b_j+k)\neq B_j(a_i-l-1)\;\;(k,l\in N_0; i=1,\ldots, n;j=1,\ldots,m),
\end{equation}
where we employ the usual notations: $N_0=(0,1,2,\ldots);R=(-\infty,\infty),R_+=(0,\infty)$ and $C$ being the complex number field. $\Omega$ is a  suitable contour separating the poles of $\Gamma(b_j+sB_j)$ from those of $\Gamma(1-a_j-A_js)$. A detailed and comprehensive account of the 
H-function is available from  Mathai and Saxena (1978).

      It follows from (7) that the generalized Mittag -Leffler function $E_{\alpha,\beta}^\gamma(z)$ can be represented in terms of the H-function in the form
     \begin{equation}
E^\gamma_{\alpha,\beta}(z)=\frac{1}{\Gamma(\gamma)}H^{1,1}_{1,2}\left[-z|^{1-\gamma,1)}_{(0,1),(1-\beta,\alpha)}\right], (Re(\alpha)>0;\alpha,\beta,\gamma\in C),
\end{equation}
and in terms of the Wright function  as
     \begin{equation}
E_{\alpha\beta}^\gamma(z)=\frac{1}{\Gamma(\gamma)}\;_1\Psi_1\left[^{(\gamma,1)}_{(\beta,\alpha)}|z\right],
\end{equation}                                                                    
where $_1\Psi_1(z)$   is a special case of  Wright's generalized hypergeometric function  
(Wright, 1935, 1940); also see Erd\'elyi et al. (1953, Section 4.1) where the function is defined by 
\begin{equation}
_p\Psi_q\left[^{(a_1,A_1),\ldots,(a_p,A_p)}_{(b_1,B_1),\ldots,(b_q, B_q)}|z\right]=
\sum^\infty_{n=0}\frac{\prod^p_{j=1}\Gamma(a_j+nA_j)}{\prod^q_{j=1}\Gamma(b_j+nB_j)}\frac{z^n}{(n)!},
\end{equation}
where $1+\sum^q_{j=1}B_j\;\;-\;\;\sum^p_{j=1}A_j\geq 0,$ (equality only holds for appropriately bounded  $z$).
The relation connecting $_p\Psi_q(z)$  and the H-function is given by Mathai and Saxena (1978, p.11, eq. 1.7.8)
\begin{equation}
_p\Psi_q\left[^{(a_1,A_1)\ldots,(a_p,A_p)}_{(b_1,B_1),\ldots,(b_q,B_q)}|z\right]=H^{1,p}_{p,q+1}\left[-z|^{(1-a_1,A_1),\ldots,(1-a_p,A_p)}_{(0,1),(1-b_1,B_1),\ldots,(1-b_q,B_q)}\right]
\end{equation}
It is interesting to observe that for $\gamma=1,\;\;$(13) and (14)  give rise to the following results
for the   generalized  Mittag-Leffler function
\begin{eqnarray}
E_{\alpha,\beta}(z)&=&\;_1\Psi_1\left[^{(1,1)}_{(\beta,\alpha)}|z\right]\\
&=&H_{1,2}^{1,1}\left[-z|^{(0,1)}_{(0,1),(1-\beta,\alpha)}\right]
\end{eqnarray} 
where $Re(\alpha)> 0, \alpha,\beta, \in C$.

      If we further take $\beta=1$ in (17) and (18) we find that 
\begin{eqnarray}
E_\alpha(z)&=&_1\Psi_1\left[^{(1,1}_{(1, \alpha)}\left|z\right]\right.\\
&=&H^{1,1}_{1,2}\left[-z\left|^{(0,1)}_{(0,1),(0,\alpha)}\right]\right.,
\end{eqnarray}
where  $Re(\alpha)>0, \alpha \in C$.
      It is shown by Kilbas et al. (2003) that  
\begin{eqnarray}
\int^x_0 & t^{\nu-1}&(x-t)^{\mu-1}E^\sigma_{\rho,\nu}(\omega t^\rho)E^\gamma_{\rho, \mu}(\omega[x-t]^\rho)dt\\\nonumber
&=& x^{\nu+\mu-1}E^{\gamma+\sigma}_{\rho,\mu,+\nu}(\omega x^\rho),
\end{eqnarray}
where $\rho, \mu, \gamma, \sigma, \omega \in C,\;\; Re(\nu)>0, Re(\mu)>0$, which is a generalization of the well-known result of Erd\'elyi et al.(1953, p.271, eq. 6.10.15)
\begin{eqnarray}
\int^x_0 & t^{\nu-1}&(x-t)^{\mu-1} \Phi(\sigma, \nu, \omega t)\Phi(\gamma, \mu, \omega[x-t])dt\\\nonumber
& =& B(\nu, \mu)x^{\nu+\mu-1} \Phi(\gamma+\sigma, \mu+\nu; \omega x),
\end{eqnarray}
where $Re(\mu)>0, Re(\nu)>0.$,
The Laplace transform of the H-function in terms of  another  H-function is given by   
Prudnikov et al. (1989, p.355, eq. 2.25.3)
\begin{equation}
L\left\{t^{\rho-1}H^{m,n}_{p,q}\left[zt^\sigma\left|^{(a_p, A_p)}_{(b_q, B_q)}\right.\right]\right\}=s^{-\rho}H^{m,n+1}_{p+1,q}\left[zs^{-\sigma}\left|^{(1-\rho,\sigma),(a_p,A_p)}_{(b_q, B_q)}\right]\right.,
\end{equation}
where $\sigma>0, Re(s)>0, Re[\rho+\sigma ^{min}_{1 \leq j\leq m}(\frac{b_j}{B_j})]>0, |arg z|< [\pi/2]\Theta, \Theta>0$;
\begin{equation}
\Theta= \sum^n_{j=1}A_j-\sum^p_{j=n+1}A_j\;\;+\;\;\sum^m_{j=1}B_j\;\;-\sum^q_{j=m+1}B_j.
\end{equation}
From (23) it  can be easily seen  that 
\begin{equation}
L^{-1}\left\{s^{-\rho}H^{m,n}_{p,q}\left[zs^\sigma|^{(a_p, A_p)}_{(b_q, B_q)}\right]\right\}= t^{\rho-1}H^{m,n}_{p+1,q}\left[zt^{-\sigma}|^{(a_p, a_p)(\rho,\sigma)}_{(b_q,B_q)}\right],
\end{equation}
where $\sigma > 0,\;\; Re(s) > 0,\;\; Re\left[\rho + \sigma ^{\mbox{max}}_{1\leq j\leq n}\left\{\frac{1-a_j}{A_j}\right\}\right]>0,\;\; |arg z|<\frac{1}{2}\pi\Theta_1, \Theta_1>0$,
where $\Theta_1=\Theta-\sigma,\;\; \Theta$ is defined in (24).
   
      From Prudnikov et al. (1989, p.355, eq. 2.25.3.2) and Mathai and Saxena (1978, p.49), it follows that the cosine transform of the H-function is given by 
\begin{eqnarray}
\int^\infty _0 & t^{\rho-1}& cos(kt)H^{m,n}_{p,q} \left[ at^\mu\left|^{(a_p, A_p)}_{(b_q,B_q)}\right]\right.dt\\\nonumber
&=&(\pi/k^\rho)H^{n+1,m}_{q+1,p+2}\left[ \frac{k^\mu}{a}\left|^{(1-b_q, B_q),(\frac{1}{2}+\frac{\rho}{2}, \frac{\mu}{2})}_{(\rho,\mu),(1-a_p,A_p,),(\frac{1}{2}+\frac{\rho}{2},\frac{\mu}{2})}\right]\right.,
\end{eqnarray}
where $Re[\rho+\mu\;\;^{\mbox{min}}_{1\leq j \leq m}(\frac{b_j}{B_j})]>0,\;\; \mu> 0,\;\; Re[\rho+\mu\;\;^{\mbox{max}}_{1\leq j \leq n}(\frac{a_j-1}{A_j})]<1,\;\;|arg\;\; a|<(\pi \theta/2), \Theta>0, \Theta$  is defined in (24).

      The Riemann-Liouville fractional integral of order $\nu\in C$ is defined by Miller and Ross (1993, p.45; see also Srivastava and Saxena, 2001) 

\begin{equation}
_0D_t^{-\nu} f(t)=\frac{1}{\Gamma(\nu)}\int_0^t (t-u)^{\nu-1} f(u)du,
\end{equation}
where $Re(\nu)>0$. Following Samko et al. (1993, p.37) we define  the fractional derivative for $\alpha > 0$ in the form
  \begin{equation}
_0D_t^\alpha f(t)=\frac{1}{\Gamma(n-\alpha)}\frac{d^n}{dt^n}\int_0^t\frac{f(u)du}{(t-u)^{\alpha-n+1}},\;\;(n=[Re(\alpha)]+1),
\end{equation}
where $[Re(\alpha)]$ means the integral part of $Re(\alpha)$.\par
\noindent
In particular, if  $0<\alpha<1$,
\begin{equation}
_0D_t^\alpha f(t)= \frac{d}{dt}\frac{1}{\Gamma(1-\alpha)}\int_0^t\frac{f(u)du}{(t-u)^\alpha}
\end{equation}
and if $\alpha=n \in N=(1,2,\ldots),$ then
\begin{equation}
_0D_t^n f(t)= D^n f(t) (D=d/dt)
\end{equation}
is the usual derivative of order n. \par
\noindent
From Erd\'elyi et al. (1954, p.182) we have
\begin{equation}
L\left\{_0D_t^{-\nu}f(t)\right\}=s^{-\nu}F(s),
\end{equation}  
where
\begin{equation}
    F(s) = L\left\{f(t); s\right\}  = \int^\infty_0 e^{-st}f(t)dt,
\end{equation}
where $Re(s) > 0$.\\ 
The Laplace transform of the fractional derivative is given by Oldham and Spanier (1974, p.134, eq. 8.1.3; see also Srivastava and Saxena, 2001):
\begin{equation}
L[_0D_t^\alpha\,f(t)]=s^\alpha F(s)-\sum^n_{k=1}s^{k-1}\,_0D_t^{\alpha-k}f(t)|_{t=0}.
\end{equation}                   
In certain boundary-value  problems arising in the theory of viscoelasticity and in the hereditary solid mechanics, Caputo (1969) introduced the following fractional  derivative  of order  $\alpha>0$, defined by  
\begin{eqnarray}
D_t^\alpha f(t)&=&\frac{1}{\Gamma(m-\alpha)}\int^t_0 \frac{f^m(\tau)d\tau}{(t-\tau)^{\alpha+1-m}}(m-1<\alpha\leq m),
Re(\alpha)>0, m\in N\nonumber\\
&=&\frac{d^m}{dt^m}f(t),\mbox{if}\;\; \alpha=m
\end{eqnarray}
      Caputo (1969) has also shown that 
\begin{equation}
L\{D_t^\alpha f(t);s\}=s^\alpha F(s)-\sum^{m-1}_{k=0}s^{\alpha-k-1}f^k(0+), (m-1<\alpha\leq m).\end{equation}
The formula (35) is very useful in the solution of differintegral equations governing certain physical problems. 

\section{Unified Fractional Kinetic Equation} 

      If we integrate the  standard kinetic equation, we obtain 
\begin{equation}
N_i(t)-N_0=-c_i\,\,_0D_t^{-1}\,\,N_i(t),\;\; (c_i>0),
\end{equation}                                                                      
where $_0D_t^{-1}$ is the standard Riemann integral operator. In the paper of  Haubold and Mathai (2000) the number density of the species $i, N_i=N_i(t)$, is a   function of time and $N_i(t=0)=N_0$ is the number density of species $i$  at time  $t = 0$. If the index  $i$  is dropped in (36), then the solution of the fractional kinetic equation 
\begin{equation}
N(t)-N_0=-c^\nu D_t^{-\nu}N(t),
\end{equation}      
is  obtained   in a series form 
\begin{equation}
N(t)=N_0\sum^\infty_{k=0}\frac{(-c^kt^k)^\nu}{\Gamma(k\nu+1)}.
\end{equation}
    
       Applying  the  definition (1), the above equation can be  rewritten in a compact 
form  in terms of the Mittag-Leffler function
\begin{equation}
     N(t)=N_0E_\nu[-(ct)^\nu], \nu>0.
\end{equation}
\noindent
{\bf \it Theorem 1.} If $c> 0,\nu > 0$, then for the solution of the integral equation 
\begin{equation}
N(t)-N_0f(t)=-c^\nu\,_0D_t^{-\nu}N(t),
\end{equation}     
where  $f(t)$  is any integrable function on the finite  interval  $[0,b]$ , there holds the formula 
\begin{equation}
N(t)=cN_0
\int^t_0H^{1,1}_{1,2}\left[c^\nu(t-\tau)^\nu\left|^{(-\frac{1}{\nu},1)}_{(-\frac{1}{\nu},1),(0,\nu)}\right.\right]f(\tau)d\tau,
\end{equation}      
where $H^{1,1}_{1,2}(.)$ is the H-function defined by  (10).\\
{\bf\it Proof:} Applying the Laplace transform to equation (40)  and using  (31)  gives
\begin{equation}
N^*(s)=L[N(t);s]=N_0\frac{F(s)}{1+(\frac{c}{s})^\nu}.
\end{equation}
where $F(s)$ is the Laplace transform of $f(t)$;\\
Since  (Mathai and Saxena, 1978, p. 152)
\begin{equation}
\frac{s^\nu}{s^\nu+c^\nu}=\,\,\,H^{1,1}_{1,1}\left[(\frac{s}{c})^\nu|^{(1,1)}_{(1.1)}\right],
\end{equation}
then using (25), we obtain 
\begin{equation}
L^{-1}\left [H^{1,1}_{1,1}\left[(\frac{s}{c})^\nu|^{(1,1)}_{(1,1)}\right]\right]=t^{-1}H^{1,1}_{2,1}[(ct)^{-\nu}|^{(1,1),(0,\nu)}_{(1,1)}].
\end{equation}
By virtue of the following  property of the H-function (Mathai and Saxena, 1978, p.4, eq. 1.2.2) 
\begin{equation}
H^{m,n}_{p,q}\left[\frac{1}{x}|^{(a_p,A_p)}_{(b_q,B_q)}\right]=H^{n,m}_{q,p}\left[x|^{(1-b_q,B_q)}_{(1-a_p,A_p)}\right],
\end{equation}
eq. (44) becomes
\begin{eqnarray}
L^{-1}\left [H_{1,1}^{1,1}\left[(\frac{s}{c})^\nu|^{(1,1)}_{(1,1)}\right]\right]&=&t^{-1}H^{1,1}_{1,2}\left[(ct)^\nu|^{(0,1)}_{(0,1),(1,\nu)}\right]\\
&=&cH^{1.1}_{1.2}\left[(ct)^\nu\left|^{(-\frac{1}{\nu},1)}_{(-\frac{1}{\nu},1),(0,\nu)}\right.\right].
\end{eqnarray}
Eq. (47) follows from (46) if we use the formula (63).\par
\noindent
Taking the inverse Laplace transform of (42) by using (47)  and applying the convolution theorem of the Laplace transform we arrive at the desired result (41).

If we set  $f(t) = t^{\rho-1}$, we obtain  the following result  established by Saxena et al. (2002, p.283, eq. 15)\\ 
{\bf\it Corollary 1.1.} If $\nu>0, \rho>0, c>0$, then for the solution of the integral equation 
\begin{equation}
N(t)-N_0t^{\rho-1}=-c^\nu D_t^{-\nu}N(t),
\end{equation}      
there holds the formula
\begin{equation}
N(t)=N_0\Gamma(\rho)t^{\rho-1}E_{\nu,\rho}[-(ct)^\nu].
\end{equation}     
Eq. (49) can be established by expressing the H-function, occurring in (41) in terms of  an
equivalent series by means of the computable representation of the H-function (Mathai and Saxena, 1978, p.71, eq. 3.7.1.) and integrating term by term by means of the beta function formula.
 
On the other hand, if we take  $f(t) = t^{\rho-1}E^\gamma_{\nu,\mu}[-(ct)^\nu]$, then  another result recently  given by Saxena et al. (2003) is obtained:\\
{\bf\it Corollary 1.2.} If $ c> 0,\;\; \nu>0,\;\; \rho>0$, then for the solution of the integral equation 
\begin{equation}
N(t)-N_0t^{\mu-1}E^\gamma_{\nu,\mu}[-(ct)^\nu]=-c^\nu\,_0D_t^{-\nu}N(t),
\end{equation}     
there holds  the formula
\begin{equation}
      N (t) = N_0 t^{\mu-1}E^{\gamma+1}_{\nu,\mu}[-(ct)^\nu].
\end{equation}
Eq. (51) can be proved with the help of the results (13) and (21).

\section{A  Fractional Diffusion Equation}

In the following we derive  the solution of  the  fractional diffusion equation using
(52). The result is obtained  in the form of the following\\ 
{\bf\it Theorem 2.}   Consider the  fractional diffusion equation (Metzler and Klafter, 2000; Jorgenson and Lang, 2001)        
\begin{equation}
_0D_t^\nu N(x,t)-\frac{t^{-\nu}}{\Gamma(1-\nu)}\delta (x)=c^\nu\frac{\partial^2}{\partial x^2}N(x,t),
\end{equation}        
with the initial condition
\begin{equation}
_0D_t^{\nu-k}N(x,t)|_{t=0}=0,\;\; (k=1,\ldots,n),
\end{equation}      
where $n=[Re(\nu)]+1,c^\nu$ is a diffusion constant and $\delta(x)$ is Dirac's delta function. Then for the solution of (52) there exists the formula 
\begin{equation}
N(x,t)=\frac{1}{(4\pi c^\nu t^\nu)^{1/2}}H^{2,0}_{1,2}\left[\frac{|x|^2}{4c^\nu t^\nu}|^{(1-\frac{\nu}{2},\nu)}_{(0,1),(1/2,1)}\right]
\end{equation}                                     
{\bf\it Proof.}  In order to derive the solution of (52), we introduce the Laplace-Fourier 
transform  in the form
\begin{equation}
N^*(k,s)=\int^\infty_0  \int^\infty_{-\infty}  e^{-st+ikx}N(x,t)dx dt.
\end{equation} 
Applying the Fourier transform with respect to the space variable  $x$  and 
Laplace transform with respect to the time variable  $t$  and using  (53), we find that 
\begin{equation}
s^\nu N^*(k,s)-s^{\nu-1}=-c^\nu k^2N^*(k,s).
\end{equation}
Solving for  $N^*(k,s)$ gives 
\begin{equation}
N^*(k,s)=\frac{s^{\nu-1}}{s^\nu+c^\nu k^2}.
\end{equation}
To invert  equation (57), it is convenient to first invert the Laplace  transform and then 
the Fourier transform . Inverting the Laplace transform , we obtain 
\begin{equation}
      N(k,t)= E_\nu(-c^\nu k^2t^\nu),
\end{equation}
which can be expressed in terms of the H-function by using  (18) as
\begin{equation}
N(k,t)  = H_{1,2}^{1,1}\left[c^\nu k^2t^\nu|^{(0,1)}_{(0,1),(0,\nu)}\right].
\end{equation}
Using the integral
\begin{equation}
\frac{1}{2\pi}\int^\infty_{-\infty}e^{-ikx}f(k)dk=\frac{1}{\pi}\int^\infty_0f(k)cos(kx)dk,
\end{equation}
and (26) to invert the Fourier transform, we see that 
\begin{eqnarray}
N(x,t) &=& \frac{1}{\pi}\int^\infty_0 cos(kx)H^{1,1}_{1,2}\left[c^\nu k^2t^\nu|^{(0,1)}_{(0,1),(0,\nu)}\right]dk\\\nonumber                                        
&=&\frac{1}{|x|}H^{2,1}_{3,3}\left[\frac{|x|^2}{c^\nu t^\nu}|^{(1,1),(1,\nu),(1,1)}_{(1,2),(1,1),(1,1)}\right].
\end{eqnarray}
Applying a result of Mathai and Saxena (1978, p.4, eq. 1.2.1) the above expression becomes
\begin{equation}
N( x,t)  = \frac{1}{|x|}H^{2,0}_{2,2}\left[\frac{|x|^2}{c^\nu t^\nu}|^{(1,\nu),(1,1)}_{(1,2),(1,1)}\right].                                       
\end{equation}
If we employ the formula (Mathai and Saxena, 1978,p. 4, eq. 1.2.4):
\begin{equation}
x^\sigma H^{m,n}_{p,q}\left[x\left|^{(a_p,A_p)}_{(b_q,B_q)}\right]\right.=H^{m,n}_{p,q}\left[x\left|^{(a_p+\sigma A_p,A_p)}_{(b_q+\sigma B_q,B_q)}\right]\right..
\end{equation}                                                 
Eq. (62) reduces to 
\begin{equation}
N(x,t)  =  \frac{1}{(c^\nu t^\nu)^{1/2}}H^{2,0}_{2,2}\left[\frac{|x|^2}{c^\nu t^\nu}|^{(1-\frac{\nu}{2},\nu),(1/2,1)}_{(0,2),(1/2,1)}\right].
\end{equation}
In view of the identity in Mathai and Saxena (1978, eq. 1.2.1), it yields
\begin{equation}
N(x,t)=\frac{1}{(c^\nu t^\nu)^{1/2}}\,H^{1,0}_{1,1}\,\left[\frac{|x|^2}{c^\nu t^\nu}\left|^{(1-\frac{\nu}{2},\nu)}_{(0,2)}\right.\right].                                             
\end{equation}
Using the definition of the H-function (10), it is seen that         
\begin{equation}
N(x,t)  =\frac{1}{2\pi\omega(c^\nu t^\nu)^{1/2}}\int_\Omega \frac{\Gamma(2\xi)}{\Gamma[1-\frac{\nu}{2}+\nu\xi]}\left[\frac{|x|^2}{c^\nu t^\nu}\right]^{-\xi}d\xi.
\end{equation}
\noindent
Applying the well-known duplication formula for the gamma function and interpreting the result thus obtained in terms of the H-function, we obtain the solution as
\begin{equation}
N(x,t)=\frac{1}{\sqrt{4\pi c^\nu t^\nu}}H^{2,0}_{1,2}\left[\frac{|x|^2}{4c^\nu t^\nu}\left|^{(1-\frac{\nu}{2},\nu)}_{(0,1),(1/2,1)}\right.\right].
\end{equation}
Finally the application of the result (Mathai and Saxena (1978, p.10, eq. 1.6.3) gives the asymptotic estimate
\begin{equation}      
N(x,t)\sim O \left\{\left[|x|^{\frac{\nu}{2-\nu}}\right][\mbox{exp}\left\{-\frac{(2-\nu)(|x|^2\nu^\nu)^{\frac{1}{2-\nu}}}{(4c^\nu t^\nu)^{\frac{1}{2-\nu}}}\right\}]\right\}\,\,(0<\nu<2).
\end{equation}
\clearpage
\begin {center}
{\bf Acknowledgment}
\end {center}
\noindent
This research is supported by a grant of  the University Grants Commission of India.    
\begin{center}
{\bf References}
\end{center}
Aslam Chaudhry, M. and Zubair, S.M.: 2002, {\it On a Class of Incomplete Gamma\par
Functions with Applications}, Chapman \& Hall/CRC, Boca Raton.\par
\noindent
Caputo, M.: 1969, {\it Elasticita e Dissipazione}, Zanichelli, Bologna.\par
\noindent
Dzherbashyan, M.M.: 1966, {\it Integral Transforms and Representation of \par
Functions in Complex Domain} (in Russian ), Nauka, Moscow.\par
\noindent
Dzherbashyan, M.M.: 1993, {\it Harmonic Analysis and Boundary Value Problems\par 
in the Complex Domain}, Birkhauser Verlag, Basel.\par
\noindent
Erd\'elyi, A.,Magnus, W., Oberhettinger,F. and Tricomi,F.G.: 1953, {\it Higher \par
Transcendental Functions}, Vol.1, McGraw-Hill, New York, Toronto,\par
and London.\par
\noindent
Erd\'elyi, A., Magnus, W. Oberhettinger, F. and Tricomi, F.G.:  1954, {\it Tables \par
of Integral Transforms}, Vol.2, McGraw-Hill, New York, Toronto,\par
and London.\par
\noindent
Erd\'elyi, A., Magnus, W., Oberhettinger, F. and Tricomi, F. G.: 1955, {\it Higher \par
Transcendental Functions}, Vol.3, McGraw- Hill, New York, Toronto, and \par
London.\par
\noindent
Haubold, H.J. and Mathai, A. M.: 1995, A heuristic remark on the periodic\par 
variation on the number of solar neutrino detected on Earth , {\it Astrophysics\par 
and Space Science} {\bf 228}, 113-134.\par
\noindent
Haubold, H.J. and Mathai, A. M.: 2000, The fractional kinetic equation and \par
thermonuclear functions, {\it Astrophysics and Space Science} {\bf 273}, 53-63.\par
\noindent
Hilfer, R. (Ed.): 2000, {\it Applications of Fractional Calculus in Physics},\par
World Scientific, Singapore.\par
\noindent
Jorgenson, J. and Lang, S.: 2001, The ubiquitous heat kernel, in {\it Mathematics\par
Unlimited - 2001 and Beyond}, Eds. B. Engquist and W. Schmid,\par
Springer-Verlag, Berlin and Heidelberg, 655-683.\par
\noindent
Kaplan, D. and Glass, L.: 1998, {\it Understanding Nonlinear Dynamics},\par
Springer-Verlag, New York, Berlin, and Heidelberg.\par
\noindent
Kilbas, A.A., Saigo, M. and Saxena, R.K.: 2002, Solution of Volterra\par
integro-differential equations with generalized Mittag-Leffler functions in\par
the kernel, {\it Journal of Integral Equations and Applications} {\bf 14},377-396.\par
\noindent
Mathai, A.M. and Saxena, R. K.: 1978, {\it The H-function with Applications in\par 
Statistics and other Disciplines}, Halsted  Press [John Wiley and Sons],\par 
New York, London and Sydney.\par
\noindent
Metzler, R. and Klafter, J.: 2000, The random walk's guide to anomalous\par
diffusion: A fractional dynamics approach, {\it Phys. Rep.} {\bf 339}, 1-77.\par
\noindent
Miller, KS and  Ross, B.: 1993, {\it An Introduction to the Fractional Calculus and \par
Fractional Differential Equations}, John Wiley and Sons, New York.\par
\noindent
Mittag-Leffler, G.M.: 1903, Sur la nouvelle fonction $E_\alpha(x)$, {\it C.R. Acad.Sci.,\par 
Paris (Ser.II)}, {\bf 137}, 554-558.\par
\noindent 
Mittag-Leffler, G.M.: 1905, Sur la representation analytique d'une fonction \par
branche uniforme d'une fonction, {\it Acta Math.} {\bf 29}, 101-181.\par
\noindent
Oldham, K.B and Spanier, J.: 1974, {\it The fractional Calculus. Theory and\par
Applications of Differentiation and Integration of Arbitrary Order},\par 
Academic Press, New York.\par
\noindent
Prabhakar, T. R.: 1971, A singular integral equation with the generalized \par
Mittag-Leffler function in the kernel, {\it Yokohama Math. J.} {\bf 19}, 7-15.\par
\noindent
Prudnikov, A.P., Brychkov, Yu.A. and Marichev, O.I.: 1989, {\it Integrals and\par 
Series. Vol.3, More Special Functions}, Gordon and Breach, New York.\par
\noindent
Samko, S.G., Kilbas, A. A. and Marichev, O.I.: 1993, {\it Fractional Integrals and \par
Derivatives: Theory and Applications}, Gordon and Breach, New York.\par
\noindent
Saxena, R.K., Mathai, A. M. and Haubold, H. J.: 2002, On fractional kinetic\par 
equations, {\it Astrophysics and Space Science} {\bf 282}, 281-287.\par
\noindent
Saxena, R.K., Mathai, A.M. and Haubold, H. J.:  2003, On generalized fractional\par 
kinetic equations, submitted for publication.\par
\noindent
Srivastava, H.M. and Saxena, R.K.: 2001, Operators of fractional integration\par
and their applications, {\it Appl. Math. Comput.} {\bf 118}, 1-52.\par
\noindent
Wright, E.M.: 1935, The asymptotic expansion of the generalized hyper-\par
geometric functions, {\it J. London Math. Soc.} {\bf 10}, 286-293.\par
\noindent
Wright, E.M.: 1940 The asymptotic expansion of the generalized hyper-\par
geometric functions, {\it Proc.London Math.Soc.} {\bf 46(2)}, 389-408.\par
\noindent
Wiman, A.:  1905, \"{U}ber den Fundamentalsatz in der Theorie der Functionen\par
$E_\alpha(x)$ {\it Acta. Math.} {\bf 29}, 191-201.
\end{document}